\documentclass[twocolumn,showpacs,preprintnumbers,pra,hyperref]{revtex4}
\usepackage{amssymb}
\usepackage{amsfonts}
\usepackage{amsmath}
\usepackage{graphicx}

\begin{document}

\title{Quantum mechanical photon-count formula derived by entangled state
representation}
\author{Li-yun Hu$^{1}$, Z. S. Wang$^{1}$, L. C. Kwek$^{2}$, and Hong-yi Fan$%
^{3}$}
\affiliation{$^{1}${\small College of Physics \& Communication Electronics, Jiangxi
Normal University, Nanchang 330022, China}\\
$^{2}${\small Center for Quantum Technologies, National University of
Singapore, Singapore 117543}\\
$^{3}${\small Department of Physics, Shanghai Jiao Tong University,
Shanghai, 200030, China}}

\begin{abstract}
{\small By introducing the thermo entangled state representation, we derived
four new photocount distribution formulas for a given density operator of
light field. It is shown that these new formulas, which is convenient to
calculate the photocount , can be expressed as such integrations over
Laguree-Gaussian function with characteristic function, Wigner function,
Q-function, and P-function, respectively.}
\end{abstract}

\maketitle

In quantum optics photon counting is important for judging the nonclassical
features of light field, most measurements of the electromagnetic field are
based on the absorption of photons via the photoelectric effect. This is
true not only for used insofar as photodiodes, photomultipliers, etc., but
also for such homely devices as the photographic plate and the eye. So the
problem of photo-electric detection attracts an increasing attention of many
physicists and scientists. Expressions for the detection probability have
been presented in many works \cite{1,2}. The quantum mechanical photon
counting distribution formula was first derived by Kelley and Kleiner \cite%
{3}. As shown in Refs. \cite{3,4,5} for the single radiation mode, the
probability distribution $\mathfrak{p}\left( m,T\right) $ of registering $m$
photoelectrons in the time interval $T$ is given by

\begin{equation}
\mathfrak{p}\left( m,T\right) =\mathtt{Tr}\left\{ \mathbf{\rho \colon }\frac{%
\left( \zeta a^{\dagger }a\right) ^{m}}{m!}e^{-\zeta a^{\dagger }a}\colon
\right\} ,  \label{p1}
\end{equation}%
where $\zeta \propto T$ is called the \textit{quantum efficiency} (a
measure) of the detector, and $\mathbf{\colon \colon }$ denotes normal
ordering. $\mathbf{\rho }$ is a single-mode density operator of the light
field concerned. The aim of this Letter is to derive some other quantum
mechanical photon-count formula by introducing the thermal entangled state
representation and convert the calculations of Wigner function (WF) and the
characteristic function of density operator to an overlap between
\textquotedblleft two pure\textquotedblright states in a two-mode enlarged
Fock space, so that it is convenient to calculate the photocount when a
light field's density operator is given. In addition, this new method seems
concise\ and easy to be accepted by readers.

Recall that the thermal entangled state representation (TESR) is constructed
in the doubled Fock space \cite{6,7} based on Umezawa-Takahash thermo field
dynamics (TFD)\ \cite{8,9,10}, i.e.,%
\begin{eqnarray}
\left\vert \eta \right\rangle &=&\exp \left[ -\frac{1}{2}|\eta |^{2}+\eta
a^{\dagger }-\eta ^{\ast }\tilde{a}^{\dagger }+a^{\dagger }\tilde{a}%
^{\dagger }\right] \left\vert 0,\tilde{0}\right\rangle  \notag \\
&=&D\left( \eta \right) \left\vert \eta =0\right\rangle ,  \label{p2}
\end{eqnarray}%
\begin{eqnarray}
\left\vert \xi \right\rangle &=&\exp \left[ -\frac{1}{2}|\xi |^{2}+\xi
a^{\dagger }+\xi ^{\ast }\tilde{a}^{\dagger }-a^{\dagger }\tilde{a}^{\dagger
}\right] \left\vert 0,\tilde{0}\right\rangle  \notag \\
&=&D\left( \xi \right) \left\vert \xi =0\right\rangle ,  \label{p4}
\end{eqnarray}%
where the state vector $\left\vert \xi \right\rangle $ is conjugate to the
state $\left\vert \eta \right\rangle ,$ $D\left( \eta \right) =e^{\eta
a^{\dagger }-\eta ^{\ast }a}$ is a displacement operator, and $\tilde{a}%
^{\dagger }$ is a fictitious mode accompanying the real photon creation
operator $a^{\dagger },$ $\left\vert 0,\tilde{0}\right\rangle =\left\vert
0\right\rangle \left\vert \tilde{0}\right\rangle ,$ and $\left\vert \tilde{0}%
\right\rangle $ is annihilated by $\tilde{a}$ with the relations $\left[
\tilde{a},\tilde{a}^{\dagger }\right] =1$ and $\left[ a,\tilde{a}^{\dagger }%
\right] =0$. It is easily seen that $\left\vert \eta =0\right\rangle $ and $%
\left\vert \xi =0\right\rangle $ have the properties,
\begin{eqnarray}
\left\vert I\right\rangle &\equiv &\left\vert \eta =0\right\rangle
=e^{a^{\dagger }\tilde{a}^{\dagger }}\left\vert 0,\tilde{0}\right\rangle
=\sum_{n=0}^{\infty }\left\vert n,\tilde{n}\right\rangle ,  \label{p3} \\
\left\vert \xi =0\right\rangle &=&(-1)^{a^{\dag }a}\left\vert \eta
=0\right\rangle ,  \label{p5}
\end{eqnarray}%
where $\tilde{n}=n$, and $\tilde{n}$\ denotes the number in the fictitious
Hilbert space.

According to the TFD and Eq.(\ref{p3}), we can reform the probability
distribution $\mathfrak{p}\left( m,T\right) $ as%
\begin{eqnarray}
\mathfrak{p}\left( m,T\right) &=&\sum_{n=0}^{\infty }\left\langle
n\right\vert \mathbf{\rho \colon }\frac{\left( \zeta a^{\dagger }a\right)
^{m}}{m!}e^{-\zeta a^{\dagger }a}\colon \left\vert n\right\rangle  \notag \\
&=&\sum_{n,l=0}^{\infty }\left\langle n,\tilde{n}\right\vert \mathbf{\rho
\colon }\frac{\left( \zeta a^{\dagger }a\right) ^{m}}{m!}e^{-\zeta
a^{\dagger }a}\colon \left\vert l,\tilde{l}\right\rangle  \notag \\
&=&\frac{\zeta ^{m}}{m!}\left\langle \mathbf{\rho }\right\vert a^{\dagger
m}\left( 1-\zeta \right) ^{a^{\dagger }a}a^{m}\left\vert I\right\rangle ,
\label{p6}
\end{eqnarray}%
where in the last step, we have used the operator identity: $\exp \left(
\lambda a^{\dag }a\right) =\colon \exp \left[ \left( e^{\lambda }-1\right)
a^{\dag }a\right] \colon $. Note that the density operators $\mathbf{\rho }$(%
$a^{\dagger }$,$a)$ are defined in the real space which are commutative with
operators ($\tilde{a}^{\dagger }$,$\tilde{a})$ in the tilde space with $%
\left\vert \rho \right\rangle \equiv \rho \left\vert I\right\rangle ,$ as
well as $\left\langle \tilde{n}\right\vert \left. \tilde{l}\right\rangle
=\delta _{n,l}$ ($n=\tilde{n},l=\tilde{l}$). By using $a^{m}\left\vert
l\right\rangle =\sqrt{l!/(l-m)!}\left\vert l-m\right\rangle ,a^{\dag
m}\left\vert l\right\rangle =\sqrt{(l+m)!/l!}\left\vert l+m\right\rangle ,$
Eq.(\ref{p6}) becomes%
\begin{eqnarray}
\mathfrak{p}\left( m,T\right) &=&\frac{\zeta ^{m}}{m!}\left\langle \mathbf{%
\rho }\right\vert \sum_{l=0}^{\infty }\frac{\left( l+m\right) !}{l!}\left(
1-\zeta \right) ^{l}\left\vert l+m,\widetilde{l+m}\right\rangle  \notag \\
&=&\zeta ^{m}\left\langle \mathbf{\rho }\right\vert \sum_{l=0}^{\infty }%
\frac{\left[ \left( 1-\zeta \right) a^{\dag }\tilde{a}^{\dag }\right] ^{l}}{%
l!}\left\vert m,\tilde{m}\right\rangle  \notag \\
&=&\zeta ^{m}\left\langle \mathbf{\rho }\right\vert e^{\left( 1-\zeta
\right) a^{\dag }\tilde{a}^{\dag }}\left\vert m,\tilde{m}\right\rangle .
\label{p7}
\end{eqnarray}

In order to derive four new formulas for $\mathfrak{p}\left( m,T\right) $,
we first bridge the relation between the characteristic function (CF) and
the entangled state representation $\left\langle \eta \right\vert $.
Similarly to Eqs.(\ref{p6}), after using the TFD theory, the CF of density
operator $\rho $, $\chi _{S}\left( \lambda ,\lambda ^{\ast }\right) =\mathtt{%
tr}\left( \rho e^{\lambda a^{\dag }-\lambda ^{\ast }a}\right) ,$ can be
calculated as%
\begin{eqnarray}
\chi _{S}\left( \lambda ,\lambda ^{\ast }\right) &=&\sum_{m,n}^{\infty
}\left\langle n,\tilde{n}\right\vert \rho e^{\lambda a^{\dag }-\lambda
^{\ast }a}\left\vert m,\tilde{m}\right\rangle  \notag \\
&=&\left\langle \rho \right\vert D\left( \lambda \right) \left\vert \eta
=0\right\rangle  \notag \\
&=&\left\langle \rho \right\vert \left. \eta =\lambda \right\rangle ,
\label{p17}
\end{eqnarray}%
which is the CF formula in thermo entangled state representation, with which
the characteristic function of density operator is simplified as an overlap
between two \textquotedblleft pure states\textquotedblright\ in enlarged
Fock space, rather than using ensemble average in the system-mode space.
Thus we can then simplify the calculation of $\chi _{S}\left( \lambda
,\lambda ^{\ast }\right) $ by virtue of some important properties of the
entangled state representation $\left\langle \eta \right\vert .$

Using the expression of $\left\langle \eta \right\vert $ in Fock space,
i.e.,
\begin{equation}
\left\langle \eta \right\vert =\left\langle 0,\tilde{0}\right\vert
\sum_{m,n=0}^{\infty }i^{m+n}\frac{a^{m}\tilde{a}^{n}}{m!n!}H_{m,n}\left(
-i\eta ^{\ast },i\eta \right) e^{-\left\vert \eta \right\vert ^{2}/2},
\label{p11}
\end{equation}%
where $H_{m,n}\left( \xi ^{\ast },\xi \right) $ is the two-variable Hermite
polynomials \cite{11,12}, one finds%
\begin{equation}
\left\langle \eta \right\vert \left. m,\tilde{n}\right\rangle
=i^{m+n}H_{m,n}(-i\eta ^{\ast },i\eta )e^{-\left\vert \eta \right\vert
^{2}/2}/\sqrt{m!n!},  \label{p12}
\end{equation}%
which leads to%
\begin{eqnarray}
&&\left\langle \eta \right\vert e^{\left( 1-\zeta \right) a^{\dag }\tilde{a}%
^{\dag }}\left\vert m,\tilde{m}\right\rangle  \notag \\
&=&\sum_{n=0}^{\infty }\frac{\left( 1-\zeta \right) ^{n}}{n!}\frac{\left(
m+n\right) !}{m!}\left\langle \eta \right. \left\vert m+n,\widetilde{m+n}%
\right\rangle  \notag \\
&=&\frac{\left( -1\right) ^{m}e^{-\left\vert \eta \right\vert ^{2}/2}}{m!}%
\sum_{n=0}^{\infty }\frac{\left( \zeta -1\right) ^{n}}{n!}H_{m+n,m+n}(-i\eta
^{\ast },i\eta )  \notag \\
&=&\frac{1}{\zeta ^{m+1}}e^{-\frac{2-\zeta }{2\zeta }\left\vert \eta
\right\vert ^{2}}L_{m}\left( \frac{1}{\zeta }\left\vert \eta \right\vert
^{2}\right) ,  \label{p13}
\end{eqnarray}%
where in the last step, we have used the formula \cite{13},
\begin{eqnarray}
&&\sum_{l=0}^{\infty }\frac{\alpha ^{l}}{l!}H_{m+l,n+l}\left( x,y\right)
\notag \\
&=&\frac{e^{\frac{\alpha \allowbreak xy}{\alpha +1}}}{\left( \alpha
+1\right) ^{(m+n+2)/2}}H_{m,n}\left( \frac{x}{\sqrt{\alpha +1}},\frac{y}{%
\sqrt{\alpha +1}}\right) ,  \label{p14}
\end{eqnarray}%
and the relation between two-variable Hermite polynomials and Laguree
polynomials,
\begin{equation}
L_{m}\left( xy\right) =\frac{(-1)^{m}}{m!}H_{m,m}\left( x,y\right) .
\label{p15}
\end{equation}%
Further inserting the completeness relation of $\left\langle \eta
\right\vert ,$i.e., $\int \frac{1}{\pi }\mathtt{d}^{2}\eta \left\vert \eta
\right\rangle \left\langle \eta \right\vert =1$ (it can be proved by using
the normally ordered form of vacuum projector $\left\vert 0,\tilde{0}%
\right\rangle \left\langle 0,\tilde{0}\right\vert =\colon \exp \left(
-a^{\dagger }a-\tilde{a}^{\dagger }\tilde{a}\right) \colon $ and the
technique of integration within an ordered product (IWOP) of operators \cite%
{14,15,16}), into Eq.(\ref{p7}), we can rewrite it as%
\begin{equation}
\mathfrak{p}\left( m,T\right) =\frac{1}{\zeta }\int \frac{\mathtt{d}%
^{2}\lambda }{\pi }e^{-\frac{2-\zeta }{2\zeta }\left\vert \lambda
\right\vert ^{2}}L_{m}\left( \frac{1}{\zeta }\left\vert \lambda \right\vert
^{2}\right) \chi _{S}\left( \lambda ,\lambda ^{\ast }\right) ,  \label{p19}
\end{equation}%
which is just a new relation about the CF and the photon-count distribution.
When the characteristic function $\chi _{S}\left( \lambda ,\lambda ^{\ast
}\right) $\ of density operator for Wigner-Weyl form is known, the
photocount distribution can be calculated by using\textsf{\ }Eq.(\ref{p19}).

For instance, we first consider the single-mode coherent states $\left\vert
\beta \right\rangle $, whose CF reads
\begin{equation}
\chi _{\text{coh}}\left( \lambda ,\lambda ^{\ast }\right) =\exp \left[ -%
\frac{1}{2}\left\vert \lambda \right\vert ^{2}+\lambda \beta ^{\ast
}-\lambda ^{\ast }\beta \right] ,  \label{p22}
\end{equation}%
substituting it into Eq.(\ref{p19}) yields
\begin{eqnarray}
\mathfrak{p}\left( m,T\right) &=&\int \frac{\mathtt{d}^{2}\lambda }{\pi
\zeta }e^{-\frac{1}{\zeta }\left\vert \lambda \right\vert ^{2}+\lambda \beta
^{\ast }-\lambda ^{\ast }\beta }L_{m}\left( \frac{1}{\zeta }\left\vert
\lambda \right\vert ^{2}\right)  \notag \\
&=&\frac{\left( \zeta \bar{n}\right) ^{m}}{m!}e^{-\zeta \bar{n}},\text{\ }(%
\bar{n}=\left\langle \beta \right\vert a^{\dagger }a\left\vert \beta
\right\rangle =\left\vert \beta \right\vert ^{2}),  \label{p23}
\end{eqnarray}%
where we use the limiting expression $\lim_{x\rightarrow
0}x^{m}L_{m}(-\left\vert \alpha \right\vert ^{2}/x)=\frac{1}{m!}\left\vert
\alpha \right\vert ^{2m}$ and the following integrational formula (see
Appendix B),
\begin{eqnarray}
&&\int \frac{d^{2}\alpha }{\pi }e^{-\allowbreak B\left\vert \alpha
\right\vert ^{2}+C\alpha -C^{\ast }\alpha ^{\ast }}L_{m}\left\{ A\left\vert
\alpha \right\vert ^{2}\right\}  \notag \\
&=&\frac{\left( B-A\right) ^{m}}{B^{m+1}}e^{\frac{-CC^{\ast }}{B}%
}L_{m}\left( \frac{ACC^{\ast }/B}{A-B}\right) .  \label{p24}
\end{eqnarray}%
Eq.(\ref{p23}) is the Poisson distribution coinciding with the result in
Refs. \cite{4,5}.

As another example, we consider the single-mode squeezed vacuum state, $\exp %
\left[ r\left( a^{\dag 2}-a^{2}\right) /2\right] \left\vert 0\right\rangle ,$
whose CF reads
\begin{equation}
\chi _{sq}\left( \lambda ,\lambda ^{\ast }\right) =\exp \left[ -\frac{1}{2}%
\left\vert \lambda \right\vert ^{2}\cosh 2r+\frac{1}{4}\left( \lambda
^{2}+\lambda ^{\ast 2}\right) \sinh 2r\right] ,  \label{p25}
\end{equation}%
substituting Eq.(\ref{p25}) into (\ref{p19}), we have (Appendix C)%
\begin{eqnarray}
\mathfrak{p}\left( m,T\right) &=&\frac{\xi ^{m}\text{sech}r\tanh ^{m}r}{%
\left( V^{2}-1\right) ^{m/2}\left( 1-V^{2}\right) ^{1/2}}P_{m}\left( \frac{V%
}{\sqrt{V^{2}-1}}\right) ,  \label{p26} \\
(V &=&\left( 1-\xi \right) \tanh r),  \notag
\end{eqnarray}%
which $P_{m}\left( x\right) $ is the Legendre polynomial and Eq.(\ref{p26})
is a new result. $\allowbreak $

Next, we derive other three new formula. Notice that the characteristic
function $\chi _{S}\left( \lambda ,\lambda ^{\ast }\right) $ is related to
the Wigner function, Q-function and P-representation by the following
Fourier transforms,%
\begin{eqnarray}
\chi _{S}\left( \lambda ,\lambda ^{\ast }\right) &=&\int e^{\lambda \alpha
^{\ast }-\lambda ^{\ast }\alpha }W\left( \alpha \right) d^{2}\alpha ,
\label{p27} \\
\chi _{S}\left( \lambda ,\lambda ^{\ast }\right) &=&e^{\frac{\left\vert
\lambda \right\vert ^{2}}{2}}\int e^{\lambda \alpha ^{\ast }-\lambda ^{\ast
}\alpha }Q\left( \alpha \right) d^{2}\alpha ,  \label{p20} \\
\chi _{S}\left( \lambda ,\lambda ^{\ast }\right) &=&e^{-\frac{\left\vert
\lambda \right\vert ^{2}}{2}}\int e^{\lambda \alpha ^{\ast }-\lambda ^{\ast
}\alpha }P\left( \alpha \right) d^{2}\alpha ,  \label{p21}
\end{eqnarray}%
respectively, thus substituting Eqs.(\ref{p27})-(\ref{p21}) into (\ref{p19})
we can directly obtain%
\begin{eqnarray}
\mathfrak{p}\left( m,T\right) &=&\frac{2\left( -\zeta \right) ^{m}}{\left(
2-\zeta \right) ^{m+1}}\int d^{2}\alpha e^{-\frac{2\zeta \left\vert \alpha
\right\vert ^{2}}{2-\zeta }}L_{m}\left\{ \frac{4\left\vert \alpha
\right\vert ^{2}}{2-\zeta }\right\} W\left( \alpha \right) ,  \label{p28} \\
\mathfrak{p}\left( m,T\right) &=&\frac{\left( -\zeta \right) ^{m}}{\left(
1-\zeta \right) ^{m+1}}\int d^{2}\alpha e^{\frac{-\zeta \left\vert \alpha
\right\vert ^{2}}{1-\zeta }}L_{m}\left\{ \frac{\left\vert \alpha \right\vert
^{2}}{1-\zeta }\right\} Q\left( \alpha \right) ,  \label{p29} \\
\mathfrak{p}\left( m,T\right) &=&\frac{\zeta ^{m}}{m!}\int d^{2}\alpha
\left\vert \alpha \right\vert ^{2m}e^{-\zeta \left\vert \alpha \right\vert
^{2}}P\left( \alpha \right) ,  \label{p30}
\end{eqnarray}%
where $W\left( \alpha \right) =2\mathtt{tr}\left( \rho \Delta \left( \alpha
,\alpha ^{\ast }\right) \right) ,Q\left( \alpha \right) =\frac{1}{\pi }%
\left\langle \alpha \right\vert \rho \left\vert \alpha \right\rangle $, and
the integrational formula (\ref{p24}) is used. Eqs.(\ref{p28})-(\ref{p30})
are the new formula for evaluating photon count distribution. Therefore,%
\textsf{\ }once one of these distributions of $\mathbf{\rho }$ is known, the
photocount distribution can be calculated by using\textsf{\ }Eq.(\ref{p28})-(%
\ref{p30}),\textsf{\ }which involve the Wigner function, Q-function, and
P-representation of $\rho $, respectively. To confirm their correctness, we
still consider the coherent light field $\left\vert \beta \right\rangle
\left\langle \beta \right\vert ,$ its Wigner function, Q-function and
P-function are given by $W\left( \alpha \right) =\frac{2}{\pi }%
e^{-2\left\vert \beta -\alpha \right\vert ^{2}},$ $P\left( \alpha \right)
=\delta ^{(2)}\left( \beta -\alpha \right) $, and \ $Q\left( \alpha \right) =%
\frac{1}{\pi }e^{-\left\vert \beta -\alpha \right\vert ^{2}}$, respectively,
then according to (\ref{p28})-(\ref{p30}) and using (\ref{p24}) and the
above limiting expression $\lim_{x\rightarrow 0}x^{m}L_{m}(-\left\vert
\alpha \right\vert ^{2}/x)=\frac{1}{m!}\left\vert \alpha \right\vert ^{2m}$,
one can draw the same result as Eq.(\ref{p23}).

At last, we should mention that using Eqs. (\ref{p2})-(\ref{p5}) it is shown
that the Wigner function of a mixed state $\rho $, $W_{\rho }\left( \alpha
,\alpha ^{\ast }\right) \equiv 2\mathtt{tr}\left( \Delta \left( \alpha
,\alpha ^{\ast }\right) \rho \right) ,$ where $\Delta \left( \alpha ,\alpha
^{\ast }\right) $ is the single-mode Wigner operator \cite{17,18}, whose
explicit normally ordered form is \cite{19}
\begin{equation}
\Delta \left( \alpha ,\alpha ^{\ast }\right) =\frac{1}{\pi }\colon
e^{-2\left( a^{\dagger }-\alpha ^{\ast }\right) \left( a-\alpha \right)
}\colon =\frac{1}{\pi }D\left( 2\alpha \right) (-1)^{a^{\dagger }a},
\label{p8}
\end{equation}%
which can also be converted to a overlap between two \textquotedblleft pure
state\textquotedblright in the enlarged Fock space,
\begin{align}
W_{\rho }\left( \alpha ,\alpha ^{\ast }\right) & =\sum_{m,n}^{\infty
}\left\langle n,\tilde{n}\right\vert \Delta \left( \alpha ,\alpha ^{\ast
}\right) \rho \left\vert m,\tilde{m}\right\rangle  \notag \\
& =\frac{1}{\pi }\left\langle \eta =0\right\vert D\left( 2\alpha \right)
(-1)^{a^{\dagger }a}\left\vert \rho \right\rangle  \notag \\
& =\frac{1}{\pi }\left\langle \eta =-2\alpha \right\vert (-1)^{a^{\dagger
}a}\left\vert \rho \right\rangle  \notag \\
& =\frac{1}{\pi }\left\langle \xi =2\alpha \right\vert \left. \rho
\right\rangle ,  \label{p9}
\end{align}%
which is the Wigner function formula in thermo entangled state
representation, with which the Wigner function of density operator is
simplified as an overlap between two \textquotedblleft pure
states\textquotedblright\ in enlarged Fock space. Employing its
completeness, i.e., $\int \frac{\mathtt{d}^{2}\xi }{\pi }\left\vert \xi
\right\rangle \left\langle \xi \right\vert =1,$ one can derive these above
new formula. In addition, the expression in Eq.(\ref{p9}) can also examine
the evolution of Wigner function of density operator interacting with the
environments \cite{20}.

In summary, based on Umezawa-Takahash thermo field dynamics theory, after
introducing the thermo entangled state representation, we converted the
calculation of CF to an overlap between two \textquotedblleft pure
states\textquotedblright\ in enlarged Fock space. Then we bridge the
relation between the characteristic function and the photo-count
distribution. Once the CF\ of density operator for Wigner-Weyl form is
known, the photocount distribution can be calculated conveniently. Using the
Fourier transform relation between the CF and the distribution functions, we
further derive other three new formula so as to be convenient for
calculating photo-count distribution by using these formulas.

\bigskip

\textbf{Acknowledgements: }Work supported by a grant from the Key Programs
Foundation of Ministry of Education of China (No. 210115) and the Research
Foundation of the Education Department of Jiangxi Province of China (No.
GJJ10097).

\bigskip

\textbf{Appendix A: Derivation of sum-formula in Eq.(\ref{p14})}

Using the integration of two-variable Hermite polynomials,

\begin{equation}
H_{m,n}\left( \xi ,\eta \right) =(-1)^{n}e^{\xi \eta }\int \frac{d^{2}z}{\pi
}z^{n}z^{\ast m}e^{-\left\vert z\right\vert ^{2}+\xi z-\eta z^{\ast }},
\tag{A1}
\end{equation}%
we have%
\begin{align}
& \sum_{l=0}^{\infty }\frac{\alpha ^{l}}{l!}H_{m+l,n+l}\left( x,y\right)
\notag \\
& =\sum_{l=0}^{\infty }\frac{\alpha ^{l}}{l!}(-1)^{n+l}e^{xy}\int \frac{%
d^{2}z}{\pi }z^{n+l}z^{\ast m+l}e^{-\left\vert z\right\vert ^{2}+xz-yz^{\ast
}}  \notag \\
& =e^{xy}(-1)^{n}\int \frac{d^{2}z}{\pi }z^{n}z^{\ast m}e^{-\left( \alpha
+1\right) \left\vert z\right\vert ^{2}+xz-yz^{\ast }}.  \tag{A2}
\end{align}%
Then making scale transform and using Eq.(A1) again, Eq.(A2) can be put into
the following form
\begin{align}
& \sum_{l=0}^{\infty }\frac{\alpha ^{l}}{l!}H_{m+l,n+l}\left( x,y\right)
\notag \\
& =\frac{(-1)^{n}e^{xy}}{\left( \alpha +1\right) ^{(m+n+2)/2}}\int \frac{%
d^{2}z}{\pi }z^{n}z^{\ast m}e^{-\left\vert z\right\vert ^{2}+\frac{xz}{\sqrt{%
\alpha +1}}-\frac{yz^{\ast }}{\sqrt{\alpha +1}}}  \notag \\
& =\text{Right hand side of Eq.(\ref{p14}).}  \tag{A3}
\end{align}

\bigskip

\textbf{Appendix B: Derivation of integration-formula in Eq.(\ref{p24})}

Using Eq.(\ref{p15}) and the generating function of the two-variable Hermite
polynomials,%
\begin{equation}
\left. \frac{\partial ^{m+n}}{\partial \tau ^{m}\partial \upsilon ^{n}}%
e^{-A\tau \upsilon +B\tau +C\upsilon }\right\vert _{\tau =\upsilon
=0}=\left( \sqrt{A}\right) ^{m+n}H_{m,n}\left( \frac{B}{\sqrt{A}},\frac{C}{%
\sqrt{A}}\right) ,  \tag{B1}
\end{equation}%
we find%
\begin{widetext}
\begin{align}
\int \frac{d^{2}\alpha }{\pi }L_{m}\left\{ A^{2}\left\vert \alpha
\right\vert ^{2}\right\} e^{-\allowbreak B^{2}\left\vert \alpha \right\vert
^{2}+C\alpha +C^{\ast }\alpha ^{\ast }}& =\int \frac{d^{2}\alpha }{\pi }%
\frac{(-1)^{m}}{m!}H_{m,m}\left\{ A\alpha ,A\alpha ^{\ast }\right\}
e^{-\allowbreak B^{2}\left\vert \alpha \right\vert ^{2}+C\alpha +C^{\ast
}\alpha ^{\ast }}  \notag \\
& =\frac{(-1)^{m}}{m!}\frac{\partial ^{2m}}{\partial t^{m}\partial t^{\prime
m}}e^{-tt^{\prime }}\int \frac{d^{2}\alpha }{\pi }\left. e^{-\allowbreak
B^{2}\left\vert \alpha \right\vert ^{2}+\left( C+At\right) \alpha +\left(
C^{\ast }+At^{\prime }\right) \alpha ^{\ast }}\right\vert _{t=t^{\prime }=0}
\notag \\
& =\frac{(-1)^{m}}{m!}\frac{\left( B^{2}-A^{2}\right) ^{m}}{B^{2\left(
m+1\right) }e^{-CC^{\ast }/B^{2}}}\frac{\partial ^{2m}}{\partial
t^{m}\partial \tau ^{m}}\left. e^{-t\tau +\frac{AC/B}{\sqrt{B^{2}-A^{2}}}%
\tau +\frac{AC^{\ast }/B}{\sqrt{B^{2}-A^{2}}}t\allowbreak }\right\vert
_{t=\tau =0},  \tag{B2}
\end{align}%
\end{widetext}
where we have used the formula%
\begin{equation}
\int \frac{d^{2}z}{\pi }e^{\zeta \left\vert z\right\vert ^{2}+\xi z+\eta
z^{\ast }}=-\frac{1}{\zeta }e^{-\frac{\xi \eta }{\zeta }},\text{Re}\left(
\zeta \right) <0  \tag{B3}
\end{equation}%
Using Eqs.(B1) and (\ref{p15}) again, one can get the integration-formula in
Eq.(\ref{p24}).

\bigskip

\textbf{Appendix C: Derivation of the result in Eq.(\ref{p26})}

In order to obtain Eq.(\ref{p26}), we first derive a new integral formula,
\begin{equation}
I\equiv \int \frac{d^{2}\lambda }{\pi }L_{m}\left\{ A\left\vert \lambda
\right\vert ^{2}\right\} e^{-B\left\vert \lambda \right\vert ^{2}+C\lambda
^{2}+C\lambda ^{\ast 2}}.  \tag{C1}
\end{equation}%
Using Eqs.(\ref{p15}) and (B1), Eq.(C1) can be put into the form%
\begin{widetext}
\begin{align}
I& =\frac{(-1)^{m}}{m!}\int \frac{d^{2}\lambda }{\pi }H_{m,m}\left( \sqrt{A}%
\lambda ,\sqrt{A}\lambda ^{\ast }\right) e^{-B\left\vert \lambda \right\vert
^{2}+C\lambda ^{2}+C\lambda ^{\ast 2}}  \notag \\
& =\frac{(-1)^{m}}{m!}\frac{\partial ^{2m}}{\partial t^{m}\partial \tau ^{m}}%
e^{-\tau t}\int \frac{d^{2}\lambda }{\pi }\left. e^{-B\left\vert \lambda
\right\vert ^{2}+t\sqrt{A}\lambda +\tau \sqrt{A}\lambda ^{\ast }+C\lambda
^{2}+C\lambda ^{\ast 2}}\right\vert _{t=\tau =0}  \notag \\
& =\frac{(-1)^{m}}{m!\sqrt{B^{2}-4C^{2}}}\frac{\partial ^{2m}}{\partial
t^{m}\partial \tau ^{m}}\exp \left[ -\frac{B^{2}-4C^{2}-BA}{B^{2}-4C^{2}}%
\tau t+\frac{CA\left( \tau ^{2}+t^{2}\right) }{B^{2}-4C^{2}}\right] _{t=\tau
=0},  \tag{C2}
\end{align}%
\end{widetext}
where in the last step, we used the formula \cite{21}%
\begin{align}
& \int \frac{d^{2}z}{\pi }\exp \left( \zeta \left\vert z\right\vert ^{2}+\xi
z+\eta z^{\ast }+fz^{2}+gz^{\ast 2}\right)  \notag \\
& =\frac{1}{\sqrt{\zeta ^{2}-4fg}}\exp \left[ \frac{-\zeta \xi \eta +\xi
^{2}g+\eta ^{2}f}{\zeta ^{2}-4fg}\right] ,  \tag{C3}
\end{align}%
whose convergent condition is Re($\zeta \pm f\pm g)<0,\ $Re[$(\zeta
^{2}-4fg)/(\zeta \pm f\pm g)]<0$.

Expanding the exponential item involved in Eq.(C2), we see%
\begin{widetext}
\begin{align}
I& =\frac{(-1)^{m}}{m!\sqrt{B^{2}-4C^{2}}}\sum_{n,l,k=0}^{\infty }\frac{%
\left( -1\right) ^{k}}{n!l!k!}\left. \frac{\left( B^{2}-4C^{2}-BA\right) ^{k}%
}{\left( B^{2}-4C^{2}\right) ^{k+n+l}/\left( CA\right) ^{n+l}}\frac{\partial
^{2m}}{\partial t^{m}\partial \tau ^{m}}\tau ^{2n+k}t^{2l+k}\right\vert
_{t=\tau =0}  \notag \\
& =\frac{\left( B^{2}-4C^{2}-BA\right) ^{m}}{\left( B^{2}-4C^{2}\right)
^{m+1/2}}\sum_{l=0}^{[m/2]}\frac{m!}{2^{2l}l!l!\left( m-2l\right) !}\left(
\frac{1}{y}\right) ^{2l},  \tag{C4}
\end{align}%
\end{widetext}
where
\begin{equation}
y=\frac{B^{2}-4C^{2}-AB}{2AC}.  \tag{C5}
\end{equation}%
Recalling that newly found expression of Lagendre polynomial (it is
equivalence to the well-known Legendre polynomial's expression \cite{22}),%
\begin{equation}
x^{m}\sum_{l=0}^{[m/2]}\frac{m!}{2^{2l}l!l!\left( m-2l\right) !}\left( 1-%
\frac{1}{x^{2}}\right) ^{l}=P_{m}\left( x\right) ,  \tag{C6}
\end{equation}%
the compact form for $I$ is written as
\begin{equation}
I=\frac{\left( \left( A-B\right) ^{2}-4C^{2}\right) ^{m/2}}{\left(
B^{2}-4C^{2}\right) ^{(m+1)/2}}P_{m}\left( \frac{y}{\sqrt{y^{2}-1}}\right) ,
\tag{C7}
\end{equation}%
which is a new integration formula.

Substituting Eq.(\ref{p25}) into (\ref{p19}) we have
\begin{equation}
\mathfrak{p}\left( m,T\right) =\frac{1}{\zeta }I^{\prime },  \tag{C8}
\end{equation}%
where $I^{\prime }$ shown in Eq.(C7) characteristic of
\begin{equation}
A=\frac{1}{\zeta },B=\frac{1}{\zeta }+\sinh ^{2}r,C=\frac{1}{4}\sinh 2r,
\tag{C9}
\end{equation}%
which leads to
\begin{equation}
y=\left( 1-\zeta \right) \tanh r,  \tag{C10}
\end{equation}%
\begin{equation}
A-B=\left( A-B\right) ^{2}-4C^{2}=-\sinh ^{2}r,  \tag{C11}
\end{equation}%
\begin{equation}
B^{2}-4C^{2}=\frac{1}{\zeta ^{2}}\left[ \left( 2-\zeta \right) \zeta \sinh
^{2}r+1\right] ,  \tag{C12}
\end{equation}%
and
\begin{align}
& \frac{\left( \left( A-B\right) ^{2}-4C^{2}\right) ^{m/2}}{\left(
B^{2}-4C^{2}\right) ^{(m+1)/2}}  \notag \\
& =\frac{\zeta ^{m+1}\text{sech}r\left( -\tanh ^{2}r\right) ^{m/2}}{\left(
\left( 2-\zeta \right) \zeta \tanh ^{2}r+\text{sech}^{2}r\right) ^{(m+1)/2}}
\notag \\
& =\frac{\zeta ^{m+1}\text{sech}r\left( -\tanh ^{2}r\right) ^{m/2}}{\left(
1-y^{2}\right) ^{(m+1)/2}},  \tag{C13}
\end{align}%
so
\begin{equation}
\mathfrak{p}\left( m,T\right) =\frac{\zeta ^{m}\text{sech}r\tanh ^{m}r}{%
\left( y^{2}-1\right) ^{m/2}\left( 1-y^{2}\right) ^{1/2}}P_{m}\left( \frac{y%
}{\sqrt{y^{2}-1}}\right) ,  \tag{C14}
\end{equation}%
which is the photon-count distribution of squeezed vacuum state.

\end{document}